\documentclass[english,10pt, onecolumn, notitlepage,  superscriptaddress]{revtex4-1}
\usepackage[T1]{fontenc}
\usepackage[latin9]{inputenc}
\usepackage{amstext}
\usepackage{graphicx}

\def\7{$\;$}
\def\l{\left}
\def\r{\right}
\def\be{\begin{equation}}
\def\ee{\end{equation}}
\def\bea{\begin{eqnarray}}
\def\eea{\end{eqnarray}}
\def\f{\frac}

\def\d{{\rm d}}

\usepackage{babel}
\def\be{\begin{equation}}
\def\ee{\end{equation}}
\def\bea{\begin{eqnarray}}
\def\eea{\end{eqnarray}}

\def\f{\frac}
\def\l{\left}
\def\r{\right}

\makeatother
\usepackage{multirow}
\usepackage{array}

\usepackage{babel}

\begin{document}

\title{Non-minimal derivative coupling gravity in cosmology}

\author{Burin Gumjudpai}
\affiliation{The Institute for Fundamental Study ``The Tah Poe Academia Institute'', Naresuan University, Phitsanulok 65000, Thailand}
\affiliation{The Abdus Salam International Centre for Theoretical Physics, Strada Costiera 11-34151, Trieste, Italy}
\author{Phongsaphat Rangdee}
\affiliation{The Institute for Fundamental Study ``The Tah Poe Academia Institute'', Naresuan University, Phitsanulok 65000, Thailand}


\date{\today}
\begin{abstract}
We give a brief review of the non-minimal derivative coupling (NMDC) scalar field theory in which there
is non-minimal coupling between the scalar field derivative term and the Einstein
tensor. We assume that the expansion is of power-law type or super-acceleration type for small redshift.
The Lagrangian includes the NMDC term, a free kinetic term, a cosmological constant term and a barotropic matter term.
For a value of the coupling constant
that is compatible with inflation,
 we use the combined WMAP9 (WMAP9+eCMB+BAO+ $H_0$) dataset,
 the PLANCK+WP dataset, and the PLANCK $TT,TE,EE$+lowP+Lensing+ext datasets
 to find the value of the cosmological constant in the model.
Modeling the expansion with power-law gives a negative cosmological constants while
the phantom power-law (super-acceleration) expansion gives
positive cosmological constant with large error bar.
The value obtained is of the same order as in the $\Lambda$CDM model,
since at late times the NMDC effect is tiny due to small curvature.
\end{abstract}
\maketitle


\section{Introduction}
\label{Section:Introduction}
 Recently, cosmic accelerating expansion has been confirmed by astrophysical observations.
 Amongst these are supernova type Ia (SNIa) \cite{Amanullah2010, Astier:2005qq, Goldhaber:2001a, Perlmutter:1997zf, Perlmutter:1999a, Riess:1998cb, Riess:1999ar, RiessGold2004, Riess:2007a, Tonry:2003a}, large-scale structure surveys \cite{Scranton:2003, Tegmark:2004},
 cosmic microwave background (CMB) anisotropies
 \cite{Larson:2010gs, arXiv:1001.4538, CMBXRay:2014, Masi:2002hp}
 and X-ray luminosity from galaxy clusters \cite{CMBXRay:2014, Allen:2004cd, Rapetti:2005a}.
 The acceleration is responsible by an unknown energy form called dark energy \cite{Copeland:2006a, Padmanabhan:2004av, Padmanabhan:2006a} which is typically
 in form of either cosmological constant or scalar field  \cite{Copeland:2006a, Padmanabhan:2004av, Padmanabhan:2006a, AT:DE2010}.
 There are many scalar field models proposed to explain the accelerating expansion of the universe, for example,
 quintessence \cite{Caldwell:1997ii} and classes of k-essence type models \cite{Chiba:1999ka, ArmendarizPicon:2000dh, ArmendarizPicon:2000ah}.
 Modifications of gravity, for instance, braneworlds, $f(R)$ and others are as well
 possible answers of present acceleration (see e.g.  \cite{AntoShijiLRR2010, Carroll2004}). Acquiring the acceleration needs
 the effective equation of state of matter species, especially a dynamical scalar field evolving under its potential, to be $p < - \rho c^2/3$.

 It is possible to have a non-minimal coupling (NMC) between scalar field to Ricci scalar in GR in form  of $\sqrt{-g}f(\phi)R$.
 The NMC is motivated by scalar-tensor theories in the Jordan-Brans-Dicke models \cite{Brans, Dirac},
 re-normalizing term of quantum field in curved space \cite{Davies} or  supersymmetries, superstring and induced gravity theories
\cite{Zee, Cho, Salam, Accetta, Maeda}. It was applied to extended inflations with first-order phase transition and other inflationary models
 \cite{ei, fs, y, ac, fm, kasper, Amendola:1993it}.
 In context of quintessence field driving present acceleration,
 non-minimal coupling to curvature has been studied as in  \cite{q, w, e, Easson:2006jd}. In strong coupling regime, power-law and de-Sitter expansions
  are found as late time attractor  \cite{Amendola:1999qq} and moreover the NMC term could also behave as effective cosmological constant \cite{CapozzRitis}.

First cosmological consideration of the non-minimal curvature coupling to the derivative term of scalar field was proposed by Amendola in 1993 \cite{Amendola1993}.
Therein the coupling function is in form of $f(\phi, \phi_{,\mu}, \phi_{,\mu\nu}, \ldots)$.
This type of derivative coupling is required in scalar quantum electrodynamics to satisfy U(1) invariance of the theory and
is required in models of which the gravitational constant is function of the mass density of the gravitational source.
The non-minimal derivative coupling-NMDC terms are commonly found as lower energy limits of higher dimensional
theories which makes quantum gravity possible to be studied perturbatively. They are also found in Weyl anomaly in $\mathcal{N} = 4$ conformal supergravity
\cite{Liu:1998bu, Nojiri:1998dh}.
With simplest NMDC term, $R \phi_{,\mu}\phi^{,\mu}$,
class of inflationary attractors is enlarged from the previous NMC model of \cite{Amendola:1993it} and the NMDC renders
 non-scale invariant spectrum without requirement of multiple scalar fields. Moreover it is possible to realize double inflation without
 adding more fields to the theory \cite{Amendola1993}.
 However conformation transformation can not transform the NMDC theory into the standard field equation in Einstein frame.
 The conformal (metric) re-scaling transformation needs to be
 generalized to Legendre transformation in order to recover the Einstein frame equations \cite{Amendola1993, mag}.
 There are various versions of the NMDC  proposed in order to match plausible theory and to predict observation results as will be seen in the next section.

We give a brief review of the NMDC gravity models in this paper and
we consider a model in which the Einstein tensor couples to the kinetic scalar field term with a free kinetic term and a constant potential (considered as a cosmological constant).
In setups of  power-law or phantom power-law (super) acceleration expansions and using inflation-estimated value of the coupling constant, we evaluate
value of the cosmological constant and show a parametric plots of the  cosmological constant versus  the power-law exponents.
Cosmological parameters given by WMAP9
(combined WMAP9+eCMB+BAO+$H_0$) dataset
\cite{cWWMAP9:2012} and PLANCK satellite dataset \cite{PLANCK:2013, Ade:2015xua} are used here.

\section{Non-minimal derivative coupling theory}

\subsection{Capozziello, Lambiase and Schmidt's result}
 Capozziello, Lambiase and Schmidt  \cite{Capozziello:1999xt} found in 2000 that
 all other possible coupling Lagrangian terms  are not necessary in scalar-curvature coupling theory, leaving only
 $R \phi_{,\mu}\phi^{,\mu}$ and  $R^{\mu\nu} \phi_{,\mu} \phi_{,\nu}$ terms in the
 Lagrangian without losing its generality, hence motivating cosmological study in the case of having
 both terms. One character of the two new terms is to modulate gravitational strength with a free canonical kinetic term
 without either scalar field potential $V(\phi)$ or $\Lambda$.  This  results in an
 effective cosmological constant and hence effectively giving de-Sitter expansion  \cite{Capozziello:1999uwa}.
 The conditions for which de-Sitter expansion is a late time attractor are given in \cite{Capozziello:1999xt}.
 When considering only $R \phi_{,\mu}\phi^{,\mu}$ with free Ricci scalar, free kinetic term, potential and matter terms,
 the equation of state, in absence of $V(\phi)$, goes to $-1$ at late time. When assuming slowly-rolling field and power-law expansion,
 $V(\phi)$ is found directly \cite{GrandaColumbia}.
 Another case is to consider only the  $R^{\mu\nu} \phi_{,\mu} \phi_{,\nu}$ term as extra term to standard scalar field cosmology,
 i.e. a free Ricci scalar with a free kinetic scalar term and a potential, the field equation contains third-order derivatives of $\phi$
 and the continuity equation of the scalar field contains third-order derivative of $g_{\mu\nu}$.
 This model is tightly constrained in weakly coupling regime, i.e. solar system constraint puts limit of the pressure,
 $p_{\phi} < 10^{-6} \rho_{\rm c} c^2$, where $\rho_{\rm c}$ is critical density hence it can not play a role of quintessence.
If the coupling is strong with negative sign, the coupling term can flattens the slope of the inflationary potential
\cite{Daniel:2007kk}.

\subsection{Granda's two coupling constant model}

Another modification of the NMDC model is proposed by Granda in 2010 \cite{Granda:2009fh}.
The model contains the usual Einstein-Hilbert term, a scalar field kinetic term, a potential term and two separated dimensionless couplings, $\kappa$, $\eta$  re-scaled
by $1/\phi^2$ in form of $- (1/2) \kappa R \phi^{-2}  g_{\mu\nu} \phi^{,\mu}\phi^{,\nu}$ and  $- (1/2) \eta \phi^{-2}  R_{\mu \nu} \phi^{\mu}\phi^{\nu}$.
In this model when there is no free kinetic scalar term (i.e. strictly NMDC) and no potential term, NMDC term takes a role of dark matter at early stage giving the power-law dust solution,
$a \sim t^{2/3}$ for $\eta = -2\kappa$ and accelerating solution for $\eta = -\kappa -1$ where $ 0 < \kappa < 1/3$.
Acceleration at present time is assured if including the potential into the Lagrangian. Motivation of such two separated couplings comes from an attempt to
approach quantum gravity perturbatively \cite{Donoghue:1994dn}. This gives ideas of the other versions of two coupling models without
 the $1/\phi^2$ re-scaling factor \cite{Granda:2010hb, Granda:2010ex, Granda:2011zk} such as inclusion of Gauss-Bonnet invariance \cite{Granda:2011eh} or in context of Chaplygin gas
\cite{Granda:2011zy}.

\subsection{Sushkov's model}

\subsubsection{Constant or zero potential}
  Sushkov, in 2009,  \cite{Sushkov:2009}
 considered  a special case $\kappa_1 R \phi_{,\mu}\phi^{,\mu}$ and $\kappa_2 R^{\mu\nu} \phi_{,\mu} \phi_{,\nu}$  with $\kappa \equiv   \kappa_2  =  -2 \kappa_1$.
 This results in combination of the two NMDC terms into one Einstein tensor coupling to kinetic scalar field part, $\kappa  G_{\mu \nu} \phi^{,\mu}\phi^{,\nu}$.
 The chosen coupling constant $\kappa$ renders good dynamical theory, that is to say,
 the field equations contain terms with second-order derivative of $g_{\mu\nu}$ and $\phi$ at most so that
 the Lagrangian contains only divergence free tensors. Hence it consists of the $R$ term, free kinetic scalar $g_{\mu\nu} \phi^{,\mu}\phi^{,\nu}$
 and  $\kappa G_{\mu \nu} \phi^{,\mu}\phi^{,\nu}$ in absence of  $V(\phi)$.
 Cosmological study of the model for flat FLRW universe yields, for $\kappa > 0$, quasi-de-Sitter at very early stage but, for $\kappa < 0$, initial singularity at very early stage.
 For any sign of the coupling, $a \propto t^{1/3}$ at very late time  \cite{Sushkov:2009}.
 A direct modification of this model is to have a constant potential with possibility of phantom behavior of the free kinetic term \cite{Saridakis:2010mf}.
 In a range of coupling constant values, this modification enables the model to transit from de-Sitter phase to other types of
 expansions giving various fates and various origins of the universe \cite{Saridakis:2010mf}. Alternative view point from de Rham, Gabadadze and Tolley massive gravity
\cite{deRham:2010kj}, is that in the decoupling limit, the massive gravity can give rise to a theory with  $\kappa  G_{\mu \nu} \phi^{,\mu}\phi^{,\nu}$ term as a subclass of Horndeski scalar-tensor theories
\cite{deRham:2011by, Heisenberg:2014kea}.

\subsubsection{With potential but without free kinetic term}
Inspired by Sushkov's model, in case of without free kinetic term, $ (1/2)g^{\mu \nu}\phi_{, \mu} \phi_{, \nu}$, but having Einstein tensor coupling kinetic term alone (strictly NMDC),
Gao in 2010 \cite{Gao:2010vr}, found that for $V(\phi) = 0$, the scalar field behaves like dust
in absence of other matters or in presence of pressureless matter.  Its value of the equation of state parameter suggests that it could be a candidate of
dark energy and dark matter. However the model is not viable due to
superluminal sound speed. When adding more than one Einstein tensor coupling to the kinetic term \cite{Gao:2010vr},
it  was claimed not to be likely by  \cite{Germani:2010hd}. Strictly NMDC term in curvaton model can also be seen in the work by \cite{Feng:2014tka}.

\subsubsection{Purely kinetic coupling term and a matter term}
  The Sushkov's model, in absence of potential and absence of matter Lagrangians, is not able to explain phantom acceleration, i.e. no phantom crossing.
   In order to fix the purely kinetic Lagrangian to allow phantom crossing,  in 2011, Gubitosi and Linder proposed most general Lagragians
   with purely kinetic term obeying shift symmetry. These are the  $( a_1\phi_{, \mu}\phi^{,\mu}  + a_2 \nabla^2 \phi )R$ term,
   $\phi_{, \mu}\phi_{,\nu} R^{\mu\nu}$ term and
   $R^{\alpha \beta \gamma \delta} f_{\alpha \beta \gamma \delta}(\phi_{,\mu}) $ term
   where $f_{\alpha \beta \gamma \delta}$ is a function of $\phi_{,\mu}$ and a matter term
    \cite{Gubitosi:2011sg}.
    Absence of potential helps avoiding high energy quantum correction.
 Their model is at lowest possible order of Planck mass and it verifies Sushkov's action \cite{Sushkov:2009}.
 The model achieve wide range of $w$ values from stiff  ($w=1$) to phantom crossing and is possible to result in loitering cosmological constant-like phase
 before entering matter domination phase.
Sushkov's purely kinetic model with matter Lagrangian is found to be a special case of the Fab Four theory  \cite{Charmousis:2011bf}.
Only positive coupling constant of the theory could result in phantom crossing
however it also gives non-causal scalar and tensor perturbation, hence making the purely-kinetic model discarded for inflation \cite{Bruneton:2012zk}.
Investigations of this model for $V(\phi) = 0 $ in blackhole spacetime are presented in \cite{Chen:2010ru, Chen:2010qf, Ding:2010fh, Kolyvaris:2011fk}.

\subsubsection{Adding potential term with matter term}
As another way out of problem in purely kinetic model, potential is added into the theory (without matter term).
In order to have inflation, it is found that the potential needs to be less steep than quadratic potential \cite{Skugoreva:2013ooa}.
With constant potential and matter term in the model, it is able to describe transition from inflation to matter domination epoch
without reheating and later it describes the transit to late de-Sitter epoch.
The derivative coupling to curvature is strong at early time to drive inflation since the coupling constant acts as
another cosmological constant $\Lambda_{\rm NMDC}$. At late time the scalar field behaves like dark matter and the cosmological constant (or the constant potential)
together with the NMDC term (with little effect) drives the present acceleration \cite{Sushkov:2012}.
Dynamical analysis  shows that for positive potential, the positive coupling gives unbound $\dot{\phi}$ value with restricted Hubble parameter \cite{Skugoreva:2013ooa}.
Indeed when considering constant potential and positive coupling, inflationary phase is always possible and the inflation depends solely on the value of coupling constant.
During inflation, gravitational heavy particles are less produced, if having stronger NMDC couplings to the inflaton field or to the particles  \cite{Koutsoumbas:2013boa}.
Perturbations analysis and inflationary analysis of the model with a constant potential considered as a cosmological constant was performed in
 \cite{Darabi:2013caa} to confront observational data.

\subsection{Model with negative-sign NMDC}  \label{section:modelneg}
The model is related (by Germani and Kehagias in 2011 \cite{Germani:2010hd})
to natural inflation of which pseudo-Nambu-Goldstone boson slowly rolling to create inflation
as well as related to  three-form inflation \cite{Germani:2009}.
The model is related to Higgs inflation with $V(\phi)  \sim \lambda \phi^4$ which is a NMDC coupling to gravity modification at tree-level of Higgs field \cite{Germani:2010gm}.
The Lagrangian looks similar to Sushkov's action but the free kinetic term and the NMDC term have opposite sign to each other, i.e. $g^{\mu \nu} - {G^{\mu\nu}}/M^2$.
The model gives a UV-protected inflation and enhances friction of the field dynamics gravitationally
 \cite{Germani:2011ua}. The model is found to be a special case of the Fab Four theory (see, e.g. \cite{Charmousis:2011bf, Charmousis:2014zaa}).
 Inflationary scenario of the model with quadratic potential and modifications of standard reheating by the NMDC term is found by Sadjadi and Goodarzi in 2013
   \cite{Sadjadi:2012zp}.  Tsujikawa in 2012 showed that, due to gravitational friction produced by the NMDC, even with steep potentials,
   a class of inflationary potentials is compatible with observation \cite{Tsujikawa:2012mk}.
   Particle production of this action after inflation is reported in
  \cite{Ema:2015oaa} and one slow roll parameter is necessary for describing inflation \cite{Ghalee:2014bta}.
  The NMDC coupling contributes to high-field friction making the energy scale reduce to sub-Planckian therefore more consistent to observation \cite{Yang:2015pga}.
The model is also investigated without free kinetic term for inflation \cite{Myung:2015tga}.
  As dark energy, this model with matter term and a power-law potential is possible to give phantom crossing \cite{Sadjadi:2010bz}.
  Power-law quintessence potential $V_0 \phi^n$  gives rise to oscillatory dark energy. The oscillatory NMDC quintessence satisfies
  EoS observational value  for $n < 2$ \cite{Sadjadi:2013psa, Sadjadi:2013uza} however inconsistencies are also reported in \cite{Jinno:2013fka}.
 Applying exponential and power-law potentials, perturbation analysis with combined SN Ia, BAO and CMB shows that NMDC coupling term
 has very small effect on late acceleration if it is needed to satisfy instability avoidance. This suggests that the coupling needs to be small,
 making  $9 \kappa H^2$ term in the Friedmann equation small.  Hence it behaves like quintessence at late time as it is driven by the potential.
 However at early time the NMDC coupling plays major role in driving the acceleration due to large $H$ value at inflation
\cite{Dent:2013awa}.  Phase space analysis for the case of exponential potential was performed in \cite{Huang:2014awa}. Static black hole scalar field solution of the model is found to be time dependent
\cite{Babichev:2013cya}.  Other investigation of the model on black holes and neutron stars can be seen in, e.g. \cite{Cisterna:2014nua, Cisterna:2015uya, Cisterna:2016vdx}.

\section{Equations of motion} \label{Section:NMDC}
In this work, we consider the Sushkov's model which takes the action \cite{Sushkov:2009, Sushkov:2012},
\be\label{MainAction}
S = \int d^4x\sqrt{-g}\l[\f{R}{8\pi G} - \l(\varepsilon g_{\mu\nu} + \kappa
G_{\mu\nu}\r)\phi^{,\mu}\phi^{,\nu} - 2V(\phi)\r] + S_{\rm m},
\ee
where $R$ is the Ricci scalar, $g$ is the determinant of metric tensor
$g_{\mu\nu}$, $G$ is the universal gravitational constant, $G_{\mu\nu}$ is the
Eintein tensor, $\phi$ is the scalar field, $V(\phi)$ is the scalar field
potential, $S_{\rm m}$ is ordinary matter action, $\varepsilon$ is a
constant with values  $+1(-1)$ for canonical (and phantom)
scalar field, $\kappa > 0$ is the coupling constant as in \cite{Sushkov:2009, Sushkov:2012}. Our universe is assumed to be a
spatially flat FLRW, with the metric
\be \label{LineElement}
\d s^2 = -c^2 \d t^2 + a^2(t) \d x^2,
\ee
where $a(t)$ is the scale factor and $\d  x^2$ is Euclidian metric.
Varying the action in Eq.(\ref{MainAction}) with respect to metric tensor
$g_{\mu\nu}$ using line element in Eq. (\ref{LineElement}) we obtain
\be\label{FriedmannEqn}
3H^2 = 4\pi G\dot{\phi}^2(\varepsilon - 9\kappa H^2) + 8\pi GV(\phi) + 8\pi
G\rho_{\rm m},   \label{eq_f1}
\ee
where $H$ is the Hubble parameter and $\rho_{\rm m}$ is the energy density of
matter. The Hubble parameter is a function of time $t$ and defined in a form
$
H = H(t) = {\dot{a}(t)}/{a(t)}
$.
The acceleration equation takes the form,
\be\label{AccEqn}
2\dot{H} + 3H^2 = -4\pi G  \dot{\phi}^2 \l[\varepsilon + \kappa\l(2\dot{H} + 3H^2 +
4H\ddot{\phi}\dot{\phi}^{-1}\r)\r] + 8\pi GV(\phi) - 8\pi Gp_{\rm m},  \label{eq_f2}
\ee
where $p_{\rm m}$ is the pressure of matter. The scalar field equation is
\be\label{EoM}
\varepsilon(\ddot{\phi} + 3H\dot{\phi}) - 3\kappa(H^2\ddot{\phi} +
2H\dot{H}\dot{\phi} + 3H^3\dot{\phi}) = -V_{,\phi}   \label{eq_f3}
\ee
where $V_{,\phi} \equiv \d V/ \d \phi$. The Eqs.
(\ref{eq_f1}),  (\ref{eq_f2}) and (\ref{eq_f3}) are the dynamical system of the field equations.
We can write
\be\label{PhiDD:1st}
\ddot{\phi} = - \f{V_{,\phi}}{\varepsilon - 3\kappa H^2} - \f{3}{\varepsilon - 3\kappa
H^2}\l(\varepsilon H - 2\kappa H\dot{H} - 3\kappa H^3\r)\dot{\phi},
\ee
or
\be\label{PhiDD:2nd}
\ddot{\phi} = -3H\dot{\phi} - \f{V_{,\phi}}{\varepsilon - 3\kappa H^2} + \f{6\kappa
H\dot{H}\dot{\phi}}{\varepsilon - 3\kappa H^2}.
\ee
Subtracting Eq. (\ref{AccEqn}) with (\ref{FriedmannEqn}), we obtain
\be\label{HDot}
\dot{H} = - 4\pi G\l[ \dot{\phi}^2 \l( \varepsilon + \kappa\dot{H} - 3\kappa H^2 +
2\kappa H\ddot{\phi}\dot{\phi}^{-1} \r) + p_{\rm m} + \rho_{\rm m} \r].
\ee
From above equations, energy density and pressure of
the scalar field is found to be
\be\label{RhoPhi}
\rho_\phi = \f{1}{2}\dot{\phi}^2(\varepsilon - 9\kappa H^2) + V(\phi),
\ee
and
\be\label{PPhi}
p_\phi = \f{1}{2}\dot{\phi}^2(\varepsilon - 9\kappa H^2)\l[1 +
\f{2\kappa\dot{H}(\varepsilon + 9\kappa H^2)}{(\varepsilon - 3\kappa H^2)(\varepsilon -
9\kappa H^2)}\r] - \f{2\kappa H\dot{\phi}V_{,\phi}}{\varepsilon - 3\kappa H^2} - V(\phi).
\ee
Therefore we find the equation of state parameter as follow
\be\label{EoS}
w_\phi = \f{\f{1}{2}\dot{\phi}^2(\varepsilon - 9\kappa H^2)\l(1 +
\f{2\kappa\dot{H}(\varepsilon + 9\kappa H^2)}{(\varepsilon - 3\kappa H^2)(\varepsilon -
9\kappa H^2)}\r) - \f{2\kappa H\dot{\phi}V_{,\phi}}{\varepsilon - 3\kappa H^2} -
V(\phi)}{\f{1}{2}\dot{\phi}^2(\varepsilon - 9\kappa H^2) + V(\phi)}.
\ee
Using the Friedmann equation, the potential is found as
\be\label{PotentialPhi}
V(\phi) = \f{3H^2}{8\pi G} - \f{1}{2}(\varepsilon - 9\kappa H^2)\dot{\phi}^2 -
\rho_{\rm m},
\ee
One can check if this is correct by
substituting the scalar field potential in to
Eq.(\ref{RhoPhi}) to obtain the usual Friedmann equation,
$
\rho_\phi   +  \rho_{\rm m} = {3H^2}/{8\pi G}
$.
From Eq.(\ref{HDot}), we see that
\be\label{RhoPhiPlusPPhi}
\rho_\phi + p_\phi = \dot{\phi}^2(\varepsilon + \kappa\dot{H} - 3\kappa H^2 + 2\kappa H
\ddot{\phi}\dot{\phi}^{-1}).
\ee
Using Friedmann equation and Eq. (\ref{RhoPhiPlusPPhi}), hence Eq. (\ref{HDot})
recovers its general kinematical form,
\be
\dot{H} = - 4\pi G \l[ \l({3H^2}/{8\pi G}\r) + p_{\rm m} + p_\phi \r]
\ee
and  the equation of state parameter also recovers general kinematical form,
\be\label{EoS:General}
w_\phi(H, \dot{H}, \rho_{\rm m}) = - \f{3H^2 + 2\dot{H} + 8\pi Gp_{\rm m}}{3H^2 - 8\pi G\rho_{\rm m}}.
\ee
Taking time derivative to the Friedmann equation (\ref{FriedmannEqn}), hence
\be\label{HDot:DivFR}
\dot{H} = - \f{4\pi G}{3H}\l[-\dot{\phi}\ddot{\phi}(\varepsilon - 9\kappa H^2) +
9\kappa H\dot{H}\dot{\phi}^2 - V_{,\phi}\dot{\phi} - \dot{\rho}_{\rm m} \r].
\ee
Using the continuity equation of matter, $\dot{\rho} = - 3H\rho$, with dust matter ($w_{\rm m} =
0$) to  Eq.(\ref{PhiDD:2nd}),  Eq.(\ref{HDot:DivFR}) becomes
\be\label{HDot:DivFR2}
\dot{H} = - 4\pi G\l[    \l\{ (\varepsilon - 9\kappa H^2) - 2\kappa\dot{H}\f{(\varepsilon -
9\kappa H^2)}{(\varepsilon - 3\kappa H^2)} + 3\kappa\dot{H}\r\} \dot{\phi}^2 - \f{2\kappa
H V_{,\phi}\dot{\phi}}{\varepsilon - 3\kappa H^2} + \rho_{\rm m}\r].
\ee
Rearrange to obtain the kinetic term,
\be\label{PhiDotSquare}
\dot{\phi}^2 = \f{\f{2\kappa HV_{,\phi}\dot{\phi}}{\varepsilon - 3\kappa H^2} - \rho_{\rm m} -
 \f{\dot{H}}{4\pi G}}{(\varepsilon - 9\kappa H^2) - 2\kappa\dot{H}\l(\f{\varepsilon -
9\kappa H^2}{\varepsilon - 3\kappa H^2}\r) + 3\kappa\dot{H}}.
\ee
Considering the case with constant potential, or equivalently a cosmological constant term, $V(\phi) = \Lambda/(8 \pi G)$ in
 the system, with dust and scalar field term (both free kinetic term and the NMDC term),
the Friedmann equation can be written as
\be
H^2 = H_0^2  \l[  \Omega_{\Lambda, 0}  +  \f{\Omega_{{\rm m}, 0}}{a^3}
  +   \f{\Omega_{\phi, 0}(\varepsilon - 9 \kappa H^2)}{a^6 (\varepsilon - 3 \kappa H^2)^2}   \r]
\ee
where $\Omega$ are density parameters of each component of cosmic fluids.
The system (\ref{eq_f1}), (\ref{eq_f2})  and (\ref{eq_f3}) with $\dot{\phi} = \psi(t) $  in absence of potential and barotropic fluid is a closed
autonomous dynamical system. An interesting particular solution of this system is when $\dot{\psi}_{\rm p}  =  0 = \ddot{\phi}$
where $\psi \equiv \dot{\phi}$ hence $\psi_{\rm p} = \dot{\phi} = \text{constant}$.
As found in \cite{Capozziello:1999uwa}, that the solution is a  de-Sitter type. For the case of $\kappa \equiv   \kappa_2  =  -2 \kappa_1$, as of Sushkov's model, the solution gives,
\be
H^2  =  \f{\Lambda_{\rm NMDC}}{3}\,.
\ee
The effective cosmological constant is defined as
\be
\Lambda_{\rm NMDC}  =  \f{\varepsilon}{ \kappa}   \label{eq_tttt}
\ee
The solution is found as $
\psi_{\rm p}  =  \dot{\phi} = {1}/{\sqrt{\kappa}}
$
which is
\be
\phi_{\rm p} \,=\,  \f{t}{\sqrt{\kappa}} + \phi_{0}
\ee
suggesting that the coupling constant should take a positive value and the effective cosmological
constant, $\Lambda_{\rm NMDC} $ should be positive. However general consideration
in \cite{Sushkov:2009, Sushkov:2012, Saridakis:2010mf} the NMDC term is strong at early time hence gives new inflation mechanism
that transition from a quasi-de-Sitter phase to power-law phase happens naturally.
Having constant $V = \Lambda/(8 \pi G)$, at late time, the transition from quasi-de-Sitter to de-Sitter phase is also possible.
The particular solution suggests that $\Lambda_{\rm NMDC} > 0$. Therefore, in presence of the usual
 cosmological constant (or constant $V$), both $\Lambda$ and $\Lambda_{\rm NMDC}$
 contribute both at late time.  In order to have enough inflation, $\kappa$ is estimated to $10^{-74}\;{\sec}^{2}$.
 Although $\Lambda_{\rm NMDC} \approx 10^{74}\;{\sec}^{-2}$ seems to be large, the NMDC term is suppressed
 by its multiplication with curvature which is very small at late time.

\section{Results}
\label{Section:PLC}
We estimate that the present universe in very recent range of $z$ evolves as power-law $
a = a_0 \l({t}/{t_0}\r)^\alpha
$ for $\varepsilon = +1$.
Here $a_0$ is scale factor at a present time, $t_0$ is age of the universe and $\alpha$ is constant exponent. The power-law expansion has been considered widely in astrophysical observations, see e.g.
 \cite{Sethi2005, Kumar2011, GumjudpaiPowerlaw1, Gumjudpai2013oxa} (see also \cite{Rani:2014sia} for constraints).
 It is realized as an attractor solution of a canonical scalar field evolving under exponential potential
\cite{Lucchin} and solution of  a barotropic fluid-dominant universe.
 Space is under acceleration if  $\alpha > 1$. We consider constant  $\alpha$ in
a range $0 < \alpha < \infty$. Hence,
$ \dot{a} = {\alpha a}/{t},
$
and the acceleration is
$
\ddot{a} = {\alpha(\alpha-1)a}/{t^2}
$.
The Hubble parameter and its time derivative are
$
H= {\dot{a}}/{a} = {\alpha}/{t},
$ and $
\dot{H} = - {\alpha}/{t^2}.
$
The value of $\alpha$ can be evaluated with data from gravitational lensing
statistics \cite{Dev:2002sz}, compact radio source \cite{Jain2003},
X-ray gas mass fraction measurements of galaxy cluster \cite{Zhu:2007tm}.
Values of $\alpha$ from various observational data are listed in \cite{Gumjudpai2013oxa}.
To calculate $\alpha$ at the present we use $\alpha = H_0 t_0$ and dust
density is
$
\rho_{\rm m} = \rho_{\rm m,0}\l({t_0}/{t}\r)^{3\alpha},
$
where $\rho_{\rm m,0}$ is the dust density at present.

\begin{figure}[!h]%
\centering
\includegraphics[width=4.5in]{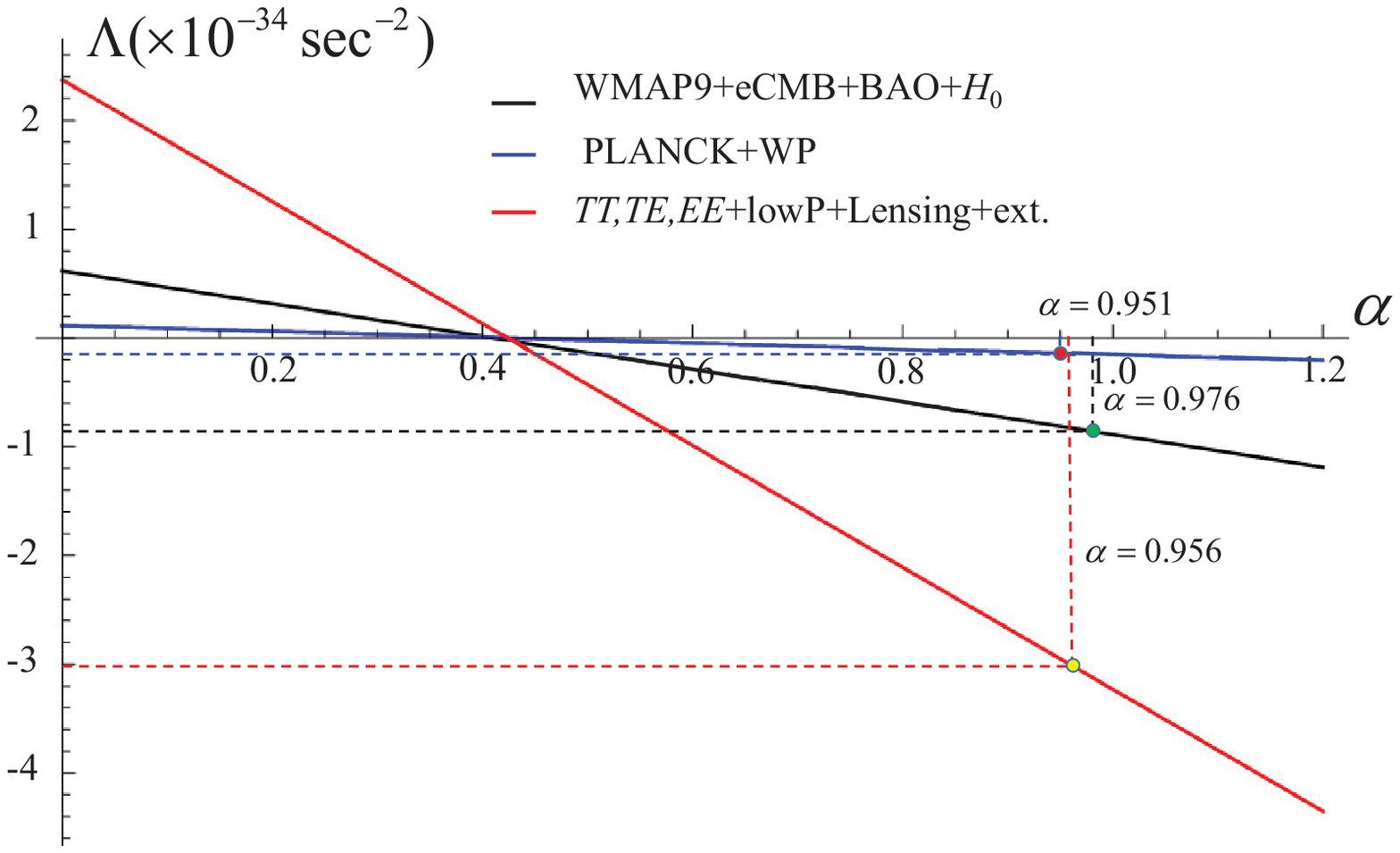}
\caption{Parametric plots of $\Lambda$ versus $\alpha$ in a power-law expansion}
\label{fig_LA}
\end{figure}

\begin{figure}[!h]%
\centering
\includegraphics[width=4.5in]{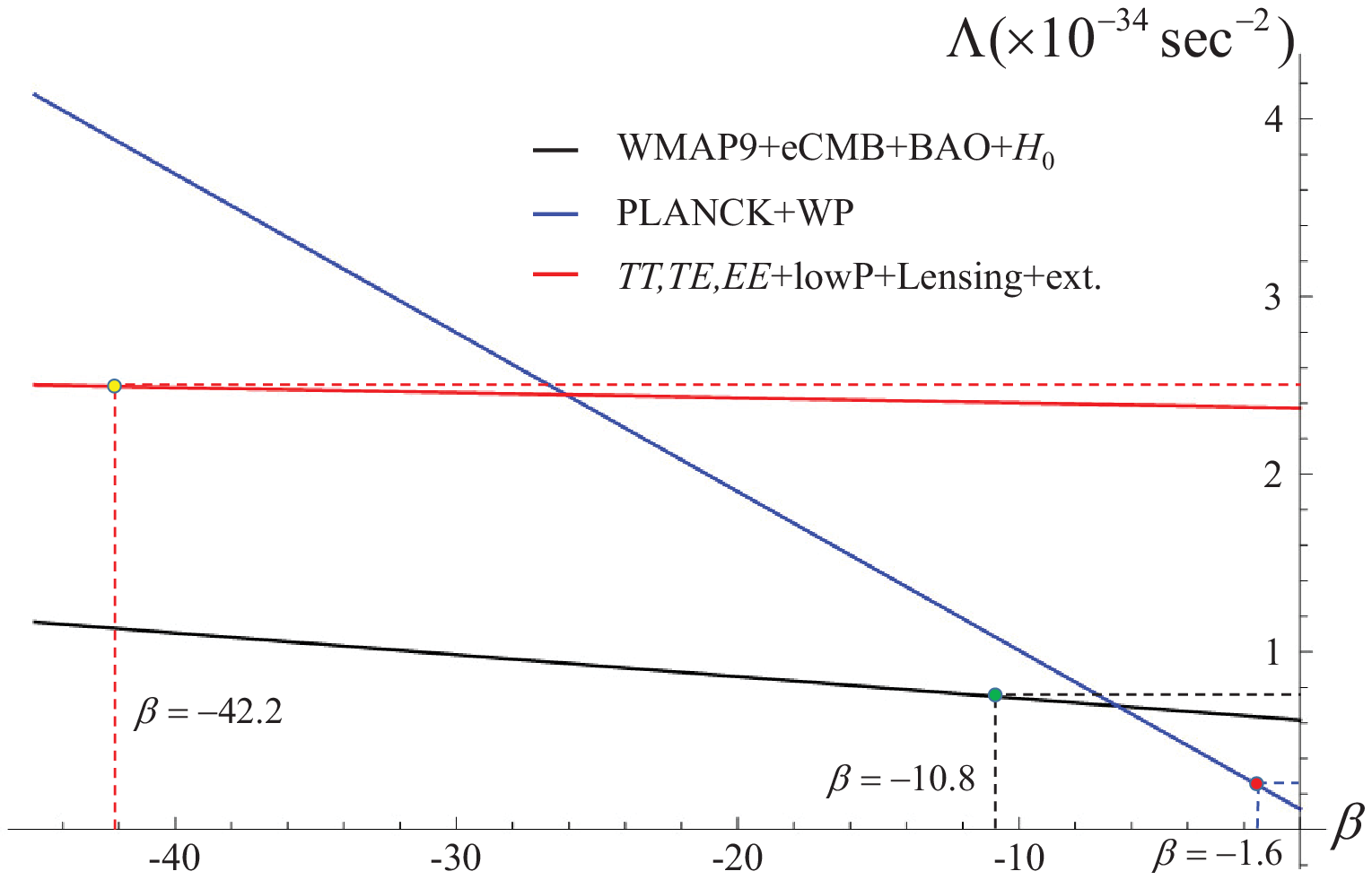}
\caption{Parametric plots of $\Lambda$ versus $\beta$ in a phantom power-law expansion}
\label{fig_LB}
\end{figure}

In the scenario of super-acceleration, i.e. the phantom power-law function for which $\varepsilon = -1$,
$
a = a_0\l[(t_{\rm s} - t)/(t_{\rm s} - t_0)\r]^\beta,
$
where $t_{\rm s}$ is the future singularity-the Big-Rip time defined as in \cite{ColesLucchin2002}
$
t_{\rm s} \equiv t_0 \,+\, {|\beta|}/{H(t_0)},
$
and $\beta$ is a constant.  In this case
$
\dot{a} = - a_0\beta {(t_{\rm s} - t)^{\beta - 1}}/{(t_{\rm s} - t_0)^\beta}
= - \beta  {a}/{(t_{\rm s} - t)},
$
and cosmic acceleration is,
$
\ddot{a} = a_0\beta(\beta - 1) {(t_{\rm s} - t)^{\beta - 2}}/{(t_{\rm s} -
t_0)^\beta} = {\beta(\beta - 1)a}/{(t_{\rm s} - t)^2}.
$
Acceleration requires $\beta < 0$. The Hubble parameter is
$
H =  - {\beta}/(t_{\rm s} - t),
$
and $
\dot{H} = - {\beta}/{(t_{\rm s} - t)^2}.
$
At present, $\beta = H_0(t_0 - t_{\rm s})$. Dust density in the
phantom power-law case is
$
\rho_{\rm m} = \rho_{\rm m,0}\l[(t_{\rm s} - t_0)/(t_{\rm s} - t)\r]^{3\beta}.
$
At present, $t = t_0$, the Big-Rip time $t_{\rm s}$ can be estimated from
\be\label{TsAtPresent}
t_{\rm s} \approx t_0 - \f{2}{3(1 + w_{\rm DE})}\f{1}{H_0\sqrt{1 - \Omega_{\rm
m,0}}}
\ee
Here, $w_{\rm DE}$ must be less than $-1$. To derive the above expression the
flat geometry and constant dark energy equation of state are assumed
\cite{Caldwell:1999ew, Caldwell:2003vq}. This type of expansion function with phantom scalar field was considered in \cite{Kaeonikhom:2010vq}.
We use
cosmological parameters are from WMAP9
(combined WMAP9+eCMB+BAO+$H_0$) dataset
\cite{cWWMAP9:2012}, PLANCK+WP  dataset \cite{PLANCK:2013} and
PLANCK including polarization and other external parameters ($TT,TE,EE$+lowP+Lensing+ext.)
\cite{Ade:2015xua}.
The value of  $w_{\rm DE}$ is of the $w$CDM model obtained from observational data. The barotropic density contributes to power-law expansion shape
while the NMDC and $\Lambda$ contributes to de-Sitter expansion, in combination, the expansion function is a mixing
between these two. For the phantom case, the free kinetic part of the Lagrangian has negative kinetic energy, therefore
the combined effect to the expansion should be the phantom-power law (super acceleration) mixing with the de-Sitter expansion.
We will calculate the cosmological constant, $\Lambda$ of the model using observed value of
 $w_{\rm DE}$  and using suggested value of $\kappa \approx 10^{-74} \;{\rm sec}^2$ as required by inflation \cite{Sushkov:2012}.
 The coupling constant is regarded as a constant in data analysis.
The derived parameters from observations are shown in Table \ref{Table:CanonicalData} while Table \ref{Table:PhantomData} shows
values of variables calculated from observations. Values of cosmological constant in this model using three datasets are shown in Table \ref{Table3}.
We show plots of $\Lambda$ versus varying value of the exponents $\alpha$ and $\beta$ in Figs. \ref{fig_LA} and \ref{fig_LB}.

\begin{table}[!h]
\begin{tabular}{|c|c|c|c|}
\hline
Parameters & WMAP9+eCMB+BAO+$H_0$ \cite{cWWMAP9:2012} & PLANCK+WP  \cite{PLANCK:2013} & \textit{TT,TE,EE}+lowP+Lensing+ext. \cite{Ade:2015xua}  \\
\hline \hline
\multirow{2}{*}{$t_0$} & $(4.346(4)\pm 0.018(6)) \times 10^{17}$ sec & $(4.360(6)\pm 0.015(1)) \times 10^{17}$ sec & $(4.354(9)\pm 0.006(6)) \times 10^{17}$ sec \\
    &   $13.772 \pm 0.059$ Gyr         &   $13.817 \pm 0.048$ Gyr & $13.799 \pm 0.021$ Gyr \\
\hline
\multirow{2}{*}{$H_0$} & $(2.245(9) \pm 0.025(9)) \times 10^{-18} {\rm sec}^{-1}$ & $(2.18(1)\pm 0.03(8)) \times 10^{-18} {\rm sec}^{-1}$ & $(2.195(1)\pm 0.014(9)) \times 10^{-18} {\rm sec}^{-1}$ \\
    &   $69.32 \pm 0.80$ km/s/Mpc                &   $67.3 \pm 1.2$ km/s/Mpc & $67.74 \pm 0.46$ km/s/Mpc \\
\hline
$\Omega_{\rm m,0}$ & $0.2865_{-0.0095}^{+0.0096}$ & $0.315_{-0.018}^{+0.016}$ & $0.3089 \pm 0.0062$ \\
\hline
$\rho_{\rm c,0}$ & $(9.019(6)\pm 0.208(8))\times 10^{-27} {\rm kg/m}^3$ & $(8.50(6)\pm 0.14(8))\times 10^{-27} {\rm kg/m}^3$ & $(8.618(6)\pm 0.117(0))\times 10^{-27} {\rm kg/m}^3$ \\
\hline
$\rho_{\rm m,0}$ & $(2.584(1)_{-0.145(5)}^{+0.146(4)})\times 10^{-27} {\rm kg/m}^3$ & $(2.67(9)_{-0.19(9)}^{+0.18(3)})\times 10^{-27} {\rm kg/m}^3$ & $(2.662(3) \pm 0.089(6))\times 10^{-27} {\rm kg/m}^3$ \\
\hline
$w_{\rm DE}$ (of $w$CDM) & $-1.073_{-0.089}^{+0.090}$ & $-1.49_{-0.57}^{+0.65}$ & $-1.019_{-0.080}^{+0.075}$ \\
\hline
\end{tabular}
\caption{Derived parameters from the combined WMAP9 (WMAP9+eCMB+BAO+$H_0$), PLANCK+WP and \textit{TT,TE,EE}+lowP+Lensing+external data.}
\label{Table:CanonicalData}
\end{table}

\begin{table}[!h]
\begin{tabular}{|c|c|c|c|}
\hline
Parameters & WMAP9+eCMB+BAO+$H_0$ & PLANCK+WP & \textit{TT,TE,EE}+lowP+Lensing+ext. \\
\hline  \hline
$\alpha$ & $0.9761(6) \pm 0.0154(3)$ & $0.951(0) \pm 0.019(9)$ & $0.9559(4) \pm 0.0079(4)$ \\
\hline
$q_{\rm power-law}$ & $0.0244(2)\pm 0.0161(9)$ & $0.0515(2)\pm 0.0220(0)$ & $0.04613(4)\pm 0.00868(9)$ \\
\hline
\multirow{2}{*}{$t_{\rm s}$} & $(5.248(1)_{-5.990(1)}^{+6.056(1)})\times 10^{18}$ sec & $(1.19(0)_{-0.91(1)}^{+1.03(2)})\times 10^{18}$ sec & $(1.96(6)_{-8.13(4)}^{+7.62(8)})\times 10^{19}$ sec \\
            &   $166.2(9)_{189.8(0)}^{191.8(9)}$ Gyr                &   $37.7(1)_{-28.8(7)}^{+32.7(0)}$ Gyr & $622.9(4)_{-2577.3(1)}^{+2416.9(8)}$ Gyr \\
\hline
$\beta$ & $-10.81(1)_{-13.58(1)}^{+13.73(1)}$ & $-1.64(4)_{-2.01(8)}^{+2.28(2)}$ & $-42.1(9)_{-178.9(9)}^{+167.8(8)}$ \\
\hline
$q_{\rm phantom}$ & $-1.0925(0)_{-0.1162(0)}^{+0.1174(8)}$ & $-1.6082(7)_{-0.7440(6)}^{+0.8443(3)}$ & $-1.0237(0)_{-0.1005(5)}^{+0.0943(1)}$ \\
\hline
\end{tabular}
\caption{Expansion derived parameters from the three datasets}
\label{Table:PhantomData}
\end{table}

\begin{table}[!h]
\begin{tabular}{|c|c|c|c|}
\hline
Parameters & WMAP9+eCMB+BAO+$H_0$ & PLANCK+WP & \textit{TT,TE,EE}+lowP+Lensing+ext. \\
\hline \hline
$\Lambda_{\rm (\varepsilon = +1)}\; ({\rm sec}^{-2})$ & $-8.5194(5)_{-4.5365(9)}^{+43.5168(8)}\times 10^{-35}$ & $-1.3997(8)_{-0.6130(8)}^{+4.5685(7)}\times 10^{-35}$ & $-2.9833(7)_{-2.3899(2)}^{+3.9590(8)}\times 10^{-34}$ \\
\hline
$\Lambda_{\rm (\varepsilon = -1)} \; ({\rm sec}^{-2})$ & $7.4792(3)_{-21.2910(5)}^{+38.2550(6)}\times 10^{-35}$ & $2.6114(3)_{-32.4699(7)}^{+8.8452(5)}\times 10^{-35}$ & $2.4939(1)_{-2.0660(4)}^{+3.3466(3)}\times 10^{-34}$  \\
\hline
\end{tabular}
\caption{Value of the cosmological  constant with power-law expansion (using $\varepsilon = +1$) and phantom power-law expansion (using $\varepsilon = -1$)
for each of observational data. \label{Table3}}
\end{table}


\section{Conclusions}
\label{Section:Conclusions}
In this work we give a brief review of the canonical scalar field model with non-minimum derivative coupling to curvature in cosmology. Of our interest in
Sushkov's model  \cite{Sushkov:2009, Sushkov:2012}, we consider the case when the potential is constant, i.e. $V = \Lambda/ (8 \pi G)$ and the coupling constant is positive.
The NMDC coupling term behaves like an effective cosmological constant, $\Lambda_{\rm NMDC} = \varepsilon/\kappa$.
Hence the NMDC term together with the free kinetic term contributes to de-Sitter like acceleration to the dynamics in the slow-roll regime at early time, i.e. inflation.
At late time the NMDC contribution is  very little due to small curvature.
At late time, in presence of  barotropic matter term and cosmological constant, we use observational data from
WMAP9+eCMB+BAO+$H_0$, PLANCK+WP and \textit{TT,TE,EE}+lowP+Lensing+external data to find cosmological constant of the theory, modeled with power-law
 and super-acceleration (phantom power-law) expansion functions. We estimate that the universe kinematically expands with power-law or super acceleration
 only from very recent redshifts.
 For  power-law expansion, the results are $\Lambda =  -8.52 \times 10^{-35}\; {\rm sec}^{-2}$ (combined WMAP9),  $-1.40\times 10^{-35}\; {\rm sec}^{-2}$ (PLANCK+WP) and
$\Lambda =  -2.98 \times 10^{-34}\; {\rm sec}^{-2}$ (\textit{TT,TE,EE}+lowP+Lensing+external data). These are of the same order as of $\Lambda$CDM model but negative.
Hence in this model, to have power-law expansion, the cosmological constant must be negative. Hence the power-law expansion is not suitable for modeling  NMDC cosmology.
For the super-acceleration  (phantom) expansion, the results are
$\Lambda =  7.48  \times 10^{-35}\; {\rm sec}^{-2}$ (combined WMAP9),
$\Lambda =  2.61  \times 10^{-35}\; {\rm sec}^{-2}$ (PLANCK+WP) and
$\Lambda =  2.49  \times 10^{-34}\; {\rm sec}^{-2}$ (\textit{TT,TE,EE}+lowP+Lensing+external data). The value is very sensitive to $t_{\rm s}$ which has large error bar.

\section*{Acknowledgments}
We thank the reviewers for useful comments. We thank Eleftherios Papantonopoulos for initiating this project and for critical comments.
B. G. is supported by National Research Council of Thailand (grant no. R2557B072).
B.G. is also sponsored via the Abdus Salam ICTP Junior Associateship scheme which supports his visit to ICTP
where this work was partially completed.
P. R. is funded via the TRF's Royal Golden Jubilee Doctoral Scholarship  (grant no. PHD/0040/2553).

\appendix
\section{Equation of state parameter for power-law case}
\label{SubSection:NMDCwCPLC}
In this part, we apply the power-law expansion $
a = a_0 \l({t}/{t_0}\r)^\alpha
$  to the NMDC cosmology. The equation of state parameter in Eq. (\ref{EoS}) takes the form
\be\label{EoS:NMDCwCPLC}
w_\phi =\f{\dot{\phi}^2(t^2 - 9\kappa\alpha^2)\l[1 - \f{2\kappa\alpha(t^2 +
9\kappa\alpha^2)}{(t^2 - 3\kappa\alpha^2)(t^2 - 9\kappa\alpha^2)}\r] -
\f{4\kappa\alpha\dot{\phi}V_{,\phi}t^3}{t^2 - 3\kappa\alpha^2} - 2V(\phi)t^2}{\dot{\phi}^2(t^2 -
3\kappa\alpha^2) + 2V(\phi)t^2}.
\ee
Eq.(\ref{PhiDotSquare}) takes the form,
\be\label{PhiDotSquare:CPLC}
\dot{\phi}^2 = \f{F_1(t, \phi, \dot{\phi})}{(t^2 - 9\kappa\alpha^2)}.
\ee
Substituting Eq.(\ref{PhiDotSquare:CPLC}) into the equation of state parameter, Eq.(\ref{EoS:NMDCwCPLC}), we obtain
\be\label{EoS:NMDCwCPLC-Full}
w_\phi = \f{ F_1(t, \phi, \dot{\phi})  \l[1 - \f{2\kappa\alpha(t^2 + 9\kappa\alpha^2)}{(t^2 - 3\kappa\alpha^2)(t^2 - 9\kappa\alpha^2)}\r] - \f{4\kappa\alpha V_{,\phi}\dot{\phi}t^3}{(t^2 - 3\kappa\alpha^2)} - 2V(\phi)t^2}{  F_1(t, \phi, \dot{\phi}) + 2V(\phi)t^2}
\ee
where
\be
F_1(t, \phi, \dot{\phi})  =  \f{\f{2\kappa\alpha V_{,\phi}\dot{\phi}t^3}{(t^2 - 9\kappa\alpha^2)} \,-\, \rho_{\rm m,0}\f{t_0^{3\alpha}}{(t^{3\alpha - 2})} \,+\, \f{\alpha}{(4\pi G)}}{1 - \f{\kappa\alpha(t^2+9\kappa\alpha^2)}{(t^2 - 3\kappa\alpha^2)(t^2 - 9\kappa\alpha^2)}}
\ee

\section{Equation of state parameter for phantom power-law case}
\label{SubSection:NMDCwPPLC}

Apply the phantom power-law expansion (super-acceleration),
$
a = a_0\l[(t_{\rm s} - t)/(t_{\rm s} - t_0)\r]^\beta,
$
The kinetic term can be written as
\be\label{PhiDotSquare:PPLC}
\dot{\phi}^2 = - \f{F_2(t, \phi, \dot{\phi})}{\l[(t_{\rm s} - t)^2 + 9\kappa\beta^2\r]}
\ee
The equation of state parameter of a phantom power-law expansion is
\be\label{EoS:NMDCwPPLC}
w_\phi = \f{     F_2(t, \phi, \dot{\phi})    \l[1 + \f{2\kappa\beta\l[(t_{\rm s} - t)^2 - 9\kappa\beta^2\r]}{\l[(t_{\rm s} - t)^2
+ 3\kappa\beta^2\r]\l[(t_{\rm s} - t)^2 + 9\kappa\beta^2\r]}\r] - \f{4\kappa\beta V_{,\phi}\dot{\phi}(t_{\rm s} - t)^3}{\l[(t_{\rm s} - t)^2 +
3\kappa\beta^2\r]} - 2V(\phi)(t_{\rm s} - t)^2}{  F_2(t, \phi, \dot{\phi})    + 2V(\phi)(t_{\rm s} - t)^2}
\ee
where
\be
F_2(t, \phi, \dot{\phi}) =    \f{\f{2\kappa\beta V_{,\phi}\dot{\phi}(t_{\rm s} - t)^3}{(t_{\rm s} - t)^2 + 3\kappa\beta^2}
- \rho_{\rm m,0}\f{(t_{\rm s} - t_0)^{3\beta}}{(t_{\rm s} - t)^{3\beta - 2}} + \f{\beta}{4\pi G}}{1 + \f{\kappa\beta\l[(t_{\rm s} - t)^2
- 9\kappa\beta^2\r]}{ \l( \l[(t_{\rm s} - t)^2 + 3\kappa\beta^2\r]\l[(t_{\rm s} - t)^2 + 9\kappa\beta^2\r] \r) } }
\ee
With constant potential in form of $
V(\phi) = {\Lambda}/{8\pi G}
$
hence $
V_{, \phi} = 0$ for both cases.

\end{document}